\documentclass{llncs}

\makeatletter
\newcommand{\@chapapp}{\relax}%
\makeatother

\usepackage{graphicx}
\usepackage{todonotes,citesort,verbatim,paralist}
\usepackage[linesnumbered,ruled,vlined]{algorithm2e}
\usepackage{ulem}
\usepackage{wrapfig}
\usepackage{appendix}
\usepackage{url}
\usepackage{cite}
\usepackage[pdfpagelabels,colorlinks,allcolors=blue]{hyperref}

\begin{document}
\title{Visualizing JIT Compiler Graphs}
\titlerunning{JIT Compiler IR Visualization}
%
\author{HeuiChan Lim \and
Stephen Kobourov}
\authorrunning{H.~Lim and S.~Kobourov}
%
\institute{Department of Computer Science, University of Arizona}
\maketitle
\begin{abstract}
Just-in-time (JIT) compilers are used by many modern programming systems in order to improve performance. Bugs in JIT compilers provide exploitable security vulnerabilities and debugging them is difficult as they are large, complex, and dynamic. Current debugging and visualization tools deal with static code and are not suitable in this domain. 
We describe a new approach for simplifying the large and complex intermediate representation, generated by a JIT compiler and visualize it with a metro map metaphor to aid developers in debugging.

\end{abstract}

\section{Introduction}\label{sec:introduction}

Many modern programming systems, such as JavaScript engines that are running our web browsers, use just-in-time (JIT) compilers to improve performance. Examples include Google Chrome, Microsoft Edge, Apple Safari, and Mozilla Firefox, which are used by 2.65 billion, 600 million, 446 million, and 220 million, respectively~\cite{safaristat}.
JIT compiler bugs can lead to exploitable security vul\-ner\-abilities~\cite{rabet2017,issue5129,issue8056,issue1072172,issue961237}. Such a bug in Google Chrome could be used to hijack passwords and to navigate to other sites and execute malicious programs, as reported by the Microsoft Offensive Security Research team  (CVE-2017-5121~\cite{rabet2017}). 
Thus, the ability to quickly analyze, localize and fix JIT compiler problems is important. However, existing work and available tools focus on static code~\cite{DBLP:conf/pldi/HolzleCU92,DBLP:conf/pldi/BrooksHS92,DBLP:conf/pldi/Adl-TabatabaiG96}, and so they are not suitable for developers in debugging  JIT compilers, which generates code at run-time. Additionally, the size and complexity of JIT-based systems~\cite{hinkelmann2017understanding} combined with the dynamic nature of JIT compiler optimizations, make it challenging to analyze and locate bugs quickly. For example, Google V8 has more than 2,000 source files and more than 1 million lines of code.

Traditional debuggers rely on text even though the main feature of a JIT compiler is building a graph-like structure to translate bytecode into optimized machine code. With this in mind, we propose a new debugging tool, which visualizes the JIT compiler's intermediate representation (IR). Our approach uses IR identification and generation techniques described by Lim and Debray~\cite{DBLP:conf/vee/LimD21}, where the compiler-related half of the visualization tool's pipeline are described in detail. In this paper we focus on the visualization half, which includes:
merging multiple IR graphs into a single graph, simplifying the merged graph, converting the simplified graph into a hypergraph, simplifying the hypergraph, and  visualizing the hypergraph using a metro map metaphor.
Visualizing the JIT compiler's IR allows us to answer questions such as:
\begin{compactenum}
\item What optimizations took place to generate the machine code?
\item What is the relationship among the optimization phases?
\item Which optimization phase was most active? 
\item What optimizations affected a specific node?
\item Which optimization phases are likely to be buggy?
\end{compactenum}

\medskip\noindent{\bf Related Work:}\label{sec:related-work} 
There are many methods and tools for debugging static code compilers and optimized code, but little on using the intermediate representation and visualizing it to show the explicit information about the compilation and optimization processes.
Google V8's Turbolizer~\cite{lukeolney2019turbolizer,jimenez2020intro} is one of very few IR visualization tools. It shows 
the final IR graph after each optimization process and provides interactive features to view the control-flow graphs for each optimization phase.  Although Turbolizer provides some information about the IR nodes and their relationships, it does not provide enough information about the optimization process and cannot answer several of our initial set of questions.

Dux {\it et al.}~\cite{DBLP:conf/iwpc/DuxIDFK05}  visualize dynamically modified code at run-time with call graphs and control-flow graphs by showing the graph changes  with animation, allowing  end-to-end play, pause, and forward/backward step-by-step animation. 
CFGExplorer~\cite{DBLP:journals/cgf/DevkotaI18} visualizes the control-flow graph of a program to represent the program structure for dynamic binary analysis. It provides interactive features allowing developers to find specific memory addresses, loops, and functions to analyze the system. CcNav~\cite{DBLP:journals/tvcg/DevkotaAKLI21}  analyzes and visualizes a C++ compiler's optimization process with a call graph, control-flow graph, and loop hierarchies. 

Control-flow graphs and call graphs are popular in program analysis, especially for analyzing  static code. However, they are different from dynamically generated IR graphs. Tools for visualizing and interacting with control-flow graphs and call graphs (such as those above) are not sufficient for visualizing the IR graph as, e.g., they cannot capture the optimization phases.

\medskip\noindent{\bf Background:}\label{sec:background}
We briefly introduce several concepts relevant to JIT compilers.

\textbf{Interpreter}:  a computer program that converts input source code into bytecode and executes it without compiling it into a machine code~\cite{DBLP:conf/hpcn/GreggEK01}.

\textbf{Bytecode}: instructions generated from input source code by an interpreter; bytecode is portable, unlike compiled programs, and used in many modern languages and systems, such as JavaScript, Python, and Java~\cite{DBLP:conf/jit/Dahm99}.

\textbf{Instruction-level Trace}: a file that holds all the instructions that a programming system, such as a JIT compiler, has generated and executed at run-time. The instructions are in a machine-level code with symbol information (e.g., function names) and are used for performance analysis and debugging.

\begin{figure}[t]
    \begin{center}
    \includegraphics[width=0.6\textwidth]{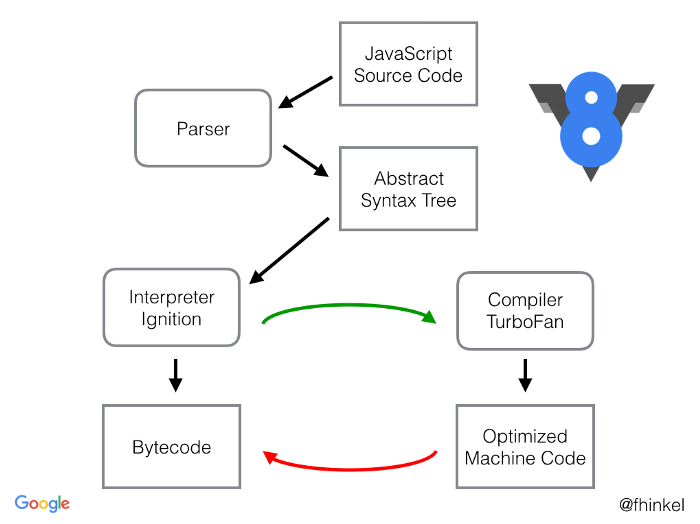}
    \includegraphics[width=0.39\textwidth]{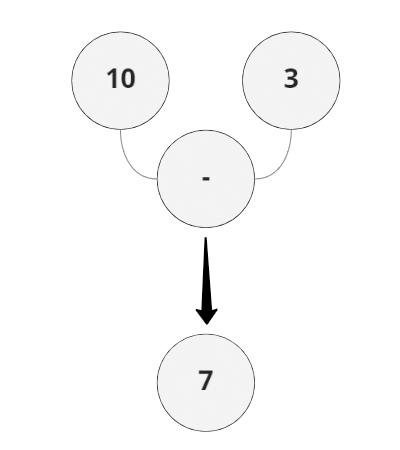}
    \end{center}
    \caption{(a) V8 Pipeline~\cite{hinkelmann2017understanding} (b) Example of constant folding optimization.}
    \label{fig:v8piepline_optexample}
\end{figure}

\textbf{Just-in-Time (JIT) compiler}: a program that turns bytecode into instructions that are sent to a computer's processor, to improve performance~\cite{DBLP:journals/concurrency/IshizakiKYTOSOKN00}; see Fig.~\ref{fig:v8piepline_optexample}(a) for an example of JIT compiler in Google's V8 pipeline.

\textbf{Optimized code}: machine code generated from bytecode by a JIT compiler that can be directly executed by a processor.

\textbf{Intermediate Representation (IR)}: a type of graph also known as sea-of-nodes~\cite{DBLP:conf/cc/DemangeRP18,meurer2016v8,Sevcik2016keyinstructions}. Unlike other graphs used in program analysis, such as control-flow or data-flow graphs which have specific types of nodes, nodes in the sea-of-nodes graph represent different types: from scalar values and arithmetic operators to variables and control-flow nodes and function entry nodes. Similarly, edges represent different relationships (e.g., semantic and syntax relationships).

\textbf{Optimization}: adding, removing, and merging nodes and edges in the graph during execution. In a single JIT compilation, the compiler executes several different optimization phases (inlining, loop peeling, constant propagation) to generate efficient machine code, which modify the IR graph and correspond to new hyperedges (the set of all nodes generated or optimized in this phase); see Fig.~\ref{fig:v8piepline_optexample}(b) for an example of constant propagation.

\textbf{Proof-of-Concept Program}: an input program that is used to trigger the buggy behavior in the JIT compiler, i.e., a valid program (without any bugs) which when run can reveal bugs in the JIT compiler. In our experiment, we are targeting JavaScript engine V8, so the PoC is a JavaScript program.

\section{Visualizing the Intermediate Representation}\label{sec:research}

Our approach for capturing and visualizing the IR of a JIT compiler below uses compiler-related steps 1-4~\cite{DBLP:conf/vee/LimD21}, and steps 5-9 are described in brief below.

\begin{compactenum}
\item Modify the input program, $P_0$, to create similar programs, $\{P_1,...P_N\}$, by generating the abstract syntax tree for $P_0$ and then randomly modifying nodes in the tree with allowable edits (passing semantic/syntactic checks). The newly created programs either still contain the code that triggers a bug in the JIT compiler, or the buggy code is replaced and no bug is triggered. In the first case, the execution output of the optimized code is different from the interpreted code (as with $P_0$).
\item Run each program $P_i$ and collect the instruction-level traces.
\item Analyze traces to check if $P_i$ triggers a bug in the JIT compiler and to identify $P_i$'s IR and the optimization phases executed while optimizing $P_i$.
\item Select candidate hyperedges, suspected to be buggy, from the information gathered in step 3.
\item Merge all selected candidate hyperedges into the original IR from $P_0$.
\item Simplify the merged IR by reducing the number of nodes and edges.
\item Convert the simplified graph into a hypergraph by extracting the hyperedges from step 4 and analyzing each node's optimization status. 
\item Simplify the hypergraph by reducing the number of hyperedges and nodes.
\item Visualize the simplified hypergraph with MetroSets~\cite{DBLP:journals/tvcg/JacobsenWKN21}.
\end{compactenum}

\subsection{Intermediate Representation}\label{sec:ir_generation}

Recall that the intermediate representation (IR) of a JIT compiler is a sea-of-nodes graph that the compiler generates at the beginning of its execution by parsing the bytecode and optimizing it with several optimization phases.
Formally, the IR is a simple, undirected graph $G = (V, E)$, where $V$ represents the nodes optimized by the JIT compiler and $E$ contains pairs of nodes connected by different relationships (e.g., semantic and syntax relationships, such as math expressions). By keeping track of the optimization information for each node we construct the hypergraph $H = (V, S)$ from $G$, where $V$ is a set of nodes optimized by the JIT compiler and each hyperedge in $S$ represents an optimization phase. 

Two important node features are phases and opcodes. Phases are the optimization phases where a node was generated and optimized (and which later correspond to hyperedges). Opcodes represent node operations (e.g., add, sub, return). A node also has two different attribute groups: (1) \textit{basic}, such as a node id, address, list of neighbors, opcode, and IR ID; and (2) \textit{optimization}, such as hyperedge (phase) ID, generated hyperedge name, and optimized hyperedge names. Note that a node is generated at one hyperedge, but can be present in multiple different hyperedges, due to different optimization phases.

Recall that given one JavaScript code we generate $N$ similar versions to see if any of them trigger bugs. We generate the IRs for all of these versions (typically about 20). In the real-world examples we work with, each such IR graph has about 300-500 nodes and 30-40 optimization phase executions.

\subsection{Merging Intermediate Representation Hyperedges}\label{sec:graph_mege}

We now merge the $N$ similar but different intermediate representations into one single graph. 
There are two main reasons to do this. First, we want to see the differences among the graphs in one single view. Second, by comparing hyperedges from a buggy program IR to hyperedges from a non-buggy program IR, we can find differences in some hyperedges due to different optimizations, and thus find the bug. Consider, for example, a hyperedge $\alpha$ in both buggy and non-buggy program IRs and suppose that an additional node (the result of incorrect optimization) makes a buggy program's $\alpha$  different from the non-buggy program's $\alpha$.  
A merged hyperedge will show this additional node, and its attributes will identify the buggy IR. 
A developer can now see that there was an optimization difference in $\alpha$ and find the bug.

\begin{figure}[t]
    \begin{center}
    \vspace{-.5cm}
    \includegraphics[width=0.49\linewidth]{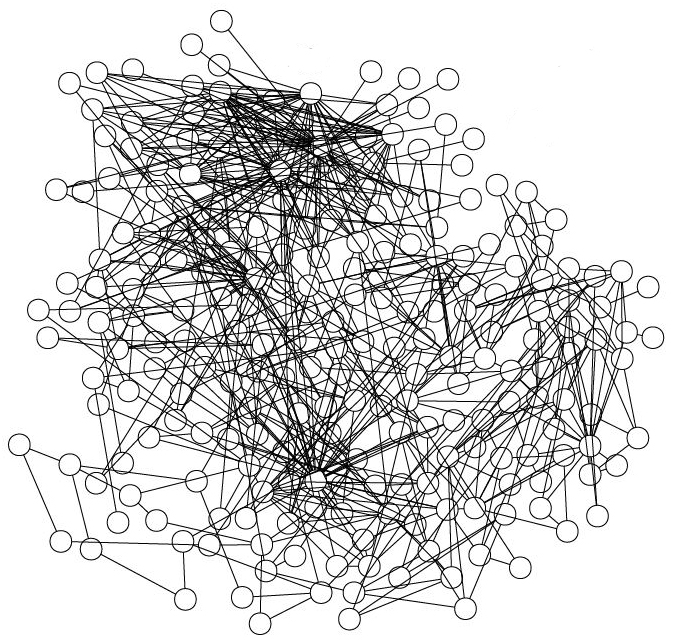}
    \includegraphics[width=0.49\linewidth]{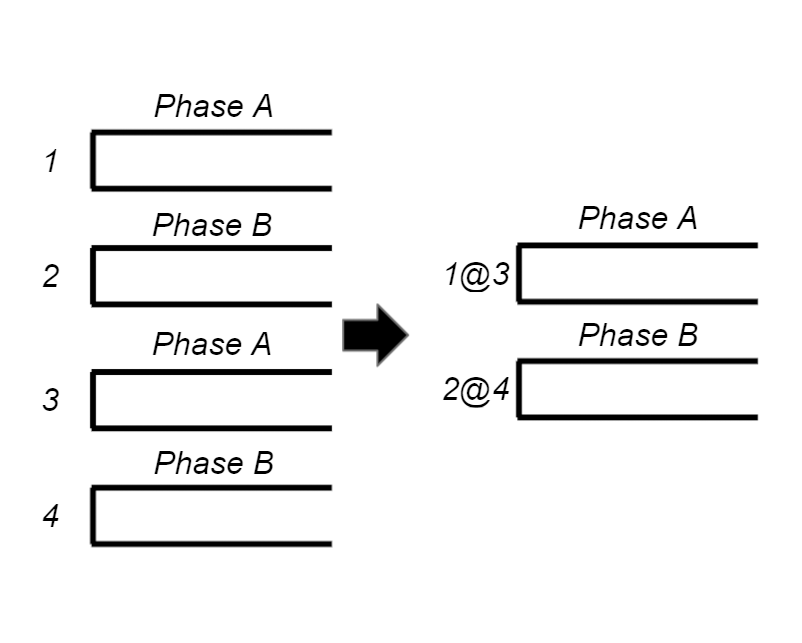}
    \vspace{-.8cm}
    \end{center}
    \caption{(a) Example of an IR graph; (b) example of hypergraph simplification.}
    \label{fig:irgraph_set_merge}
\end{figure}

Let $R_0$ be the IR from the original program and $\{R'_1,...,R'_N\}$ the IRs from the modified programs. Let $\{r'_1,...,r'_n\}$ be sub-IRs, where $r'_i$ is a subgraph of $R'_i$ when $R'_i\ne R_0$, i.e., $r'_i \subseteq R'_i$, and $n$ is the number of IRs different from $R_0$ ($n \le N$).
Each $r'_i$ holds buggy candidate hyperedges: $R'_i$ hyperedges are different from $R_0$'s hyperedges.  
We traverse all sub-IRs, comparing each to $R_0$, and update the merged IR; see Algorithm~1 in~\cite{lim2021visualizing} for detail.

\subsection{Intermediate Representation Simplification}\label{sec:graph_simplify}

Although the resulting merged graph may be useful for debugging, its complexity makes it difficult for developers to use; see  Fig.~\ref{fig:irgraph_set_merge}(a). Therefore, we simplify the graph, convert it into a hypergraph, and simplify the hypergraph (hopefully without losing much information in these simplifications). The main goal is to end up with an interactive visualization that allows developers to debug. 

\medskip\noindent{\bf Reducing the IR Graph: }\label{sec:reducegraph} 
We remove dead nodes (nodes with no adjacent edges) as they are not translated into machine code and do not affect other nodes. We then identify nodes that can be merged without losing important information. A pair of nodes is merged if they have the same opcode, the same optimization information, belong to the same IR (which can be identified by the IR id attribute), and share the same neighbors; see Algorithm~2 in~\cite{lim2021visualizing} for detail.

\medskip\noindent{\bf Reducing the IR Hypergraph:}\label{sec:reducehypergraph} We  convert the simplified graph $G=(V,E)$ into a hypergraph $H=(V,S)$, by extracting hyperedges based on the optimization phases; 
see Algorithm~3 in~\cite{lim2021visualizing}. Recall that a node $v$  generated in phase/hyperedge $\alpha$ and optimized in phases/hyperedges $\phi$ and $\gamma$ now belongs to all three hyperedges.
We reduce hypergraph $H$ by merging suitable pairs of hyperedges. Different nodes can have the same hyperedge names as attributes, but different hyperedge IDs, as IDs are assigned based on the execution order. Therefore, we merge hyperedges with the same name into a single hyperedge while assigning a new unique identifier generated from the original IDs. We use ID concatenation to obtain unique identifiers. Consider two hyperedges \textit{A} and \textit{B} executed twice in the order shown in  Fig.~\ref{fig:irgraph_set_merge}(b). We use the order to create unique IDs by merging the 4 hyperedges into 2  hyperedges and assigning new IDs, generated by concatenating two IDs delimited with a special character `@'; see Algorithm~4  in~\cite{lim2021visualizing}.

This reduces the number of hyperedges but increases the number of nodes in each hyperedge. Next, we traverse each hyperedge $s\in S$, and we use node opcodes to see if they can be merged; see Algorithm~5 and Table~1 in~\cite{lim2021visualizing} for more details and results.

\subsection{Visualizing the Hypergraph with MetroSets}\label{sec:metroset}

MetroSets~\cite{DBLP:journals/tvcg/JacobsenWKN21} 
uses the metro map metaphor to visualize medium-size hypergraphs. It clearly shows the relationships between hyperedges, which in our case captures the relationships among the optimizations. MetroSets provides simple and intuitive interactions that make it possible to quickly identify hyperedges (metro lines) that contain suspicious nodes (metro stations), or hyperedges that intersect with a particular suspicious hyperedge.
Each node in the MetroSet map is labeled with its unique ID (representing the node generation timeline). The attributes shown when hovering over a node are phase, opcode, address, graph ID, and phase ID. A phase attribute tells the user where the node was generated and it is useful when nodes belong to multiple sets. A developer can distinguish the phase that generated a node and phases where it was optimized.

\section{Evaluation}\label{sec:evaluation}

\begin{figure*}[t]
    \hspace{-.5cm}\includegraphics[width=1.1\textwidth]{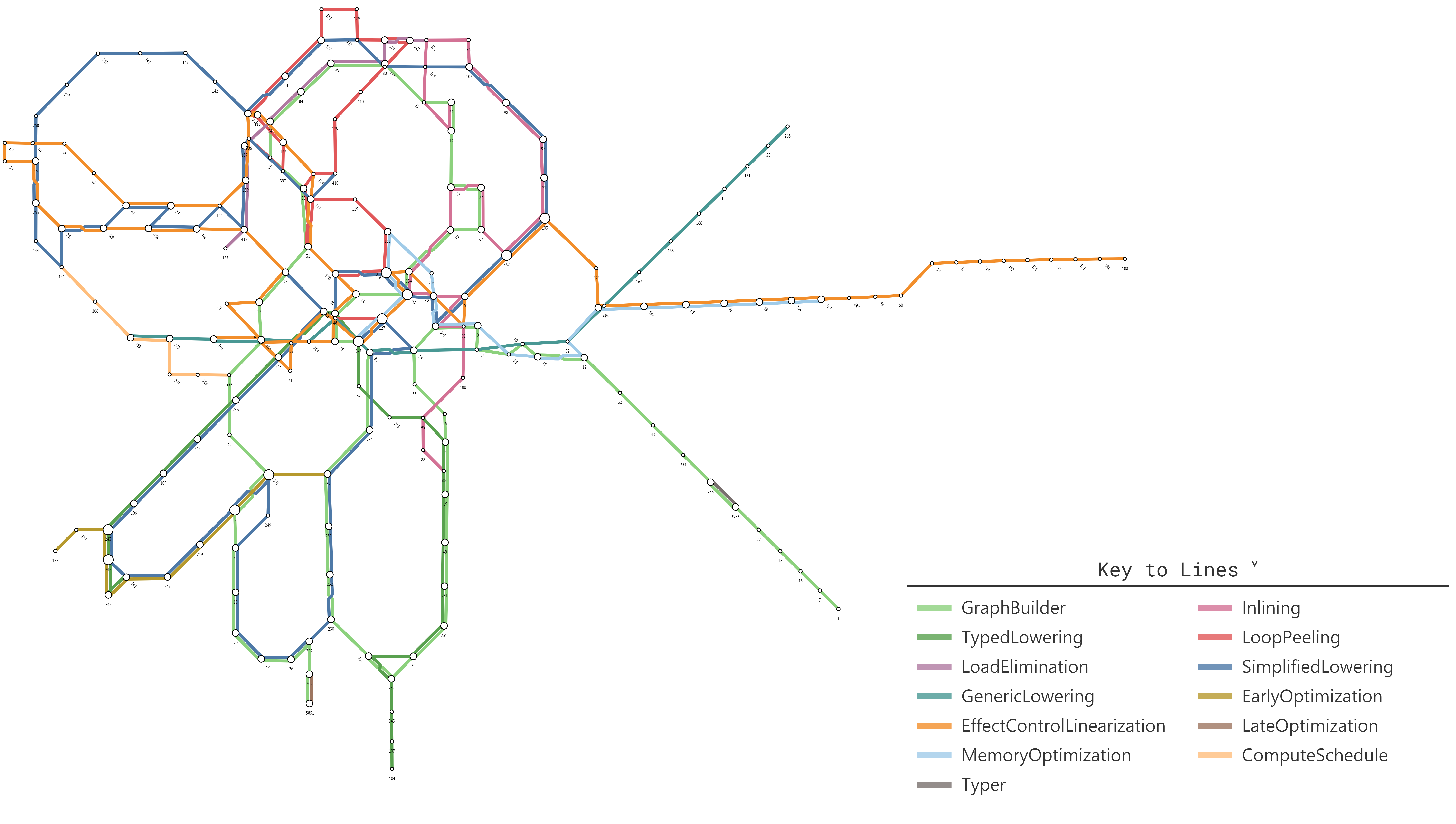}
    \caption{Metro map of the IR graph from bug report 5129.}
    \label{fig:5129_metro}
\end{figure*}

We work with Google's JavaScript engine and its JIT compiler, using a dynamic analysis tool built on top of Intel's Pin software~\cite{DBLP:conf/pldi/LukCMPKLWRH05} to collect instruction-level 
traces, XED~\cite{xed} for instruction decoding~\cite{xed},  esprima-python~\cite{esprima} to generate the syntax-tree from JavaScript code, and escodegen~\cite{escodegen} to regenerate JavaScript from the syntax-tree. 
Our data comes from the Chromium bug report site; see~\cite{DBLP:conf/vee/LimD21} for details.
We can identify the bugs in all listed bug reports, including Chromium bug report 5129. This version of the compiler has a bug in the $\it EarlyOptimization$ phase.
We generate 19 additional modified JavaScript programs from the original and run all 20.
The instruction traces are used to generate the IR graph shown in Fig.~\ref{fig:irgraph_set_merge}(a) and our visualization is shown in Fig.~\ref{fig:5129_metro}. We can now attempt to answer some of the questions from Sec.~\ref{sec:introduction}.

\textit{``What optimizations took place to generate the machine code?”} The map and the ``Key to Lines" legend show all optimization phases.

\textit{``What is the relationship among the optimization phases?”} 
We can examine the corresponding lines and use the interactive exploration modes (intersection, union, complement, etc.) to see the relationships among the phases.

\textit{``Which optimization phase was most active?”} We can visually identify the longest line, or hover over each line and see  the number of nodes in it; see Figure~9 in~\cite{lim2021visualizing} for an example of the most active optimization phase.

\textit{``What optimizations affected a specific node"} We can hover over the node of interest, which grays out the lines that don't contain the node. We can then examine each of the corresponding lines and look at the displayed node attributes.

\textit{``Which optimization phases are likely to be buggy?”} One natural way to do this is to find parts that differ in the IR graphs with the bug and those without. In other words, a program is  buggy  because either it has additional optimizations or missing optimizations, and this information is captured in the IRs. Any line that has many non-original IRs represents a significant difference between buggy and non-buggy programs. In this case study, the majority of nodes (9 out of 11) in the EarlyOptimization line are from different IRs, indicating a difference in optimization between buggy and non-buggy programs; see the full paper~\cite{lim2021visualizing} for more examples.

Our prototype is available at {\small \url{https://hlim1.github.io/JITCompilerIRViz/}}.

\subsubsection*{Acknowledgements} This research was supported in part by the National Science Foundation  under grants CNS-1908313 and DMS-1839274.

\raggedright

\clearpage

\end{document}